\documentclass[useAMS,usenatbib,epsfig]{mn2e}
\newif\ifAMStwofonts
\AMStwofontstrue
\usepackage{amsmath,amsfonts,epsfig,natbib}

\def\reff@jnl#1{{\rm#1\/}}

\def\aj{\reff@jnl{AJ}}                  
\def\araa{\reff@jnl{ARA\&A}}            
\def\apj{\reff@jnl{ApJ}}                        
\def\apjl{\reff@jnl{ApJ}}               
\def\apjs{\reff@jnl{ApJS}}              
\def\ao{\reff@jnl{Appl.Optics}}         
\def\apss{\reff@jnl{Ap\&SS}}            
\def\aap{\reff@jnl{A\&A}}               
\def\aapr{\reff@jnl{A\&A~Rev.}}         
\def\aaps{\reff@jnl{A\&AS}}             
\def\azh{\reff@jnl{AZh}}                        
\def\baas{\reff@jnl{BAAS}}              
\def\gca{\reff@jnl{GeCoA}}              
\def\jrasc{\reff@jnl{JRASC}}            
\def\memras{\reff@jnl{MmRAS}}           
\def\mnras{\reff@jnl{MNRAS}}            
\def\pra{\reff@jnl{Phys.Rev.A}}         
\def\prb{\reff@jnl{Phys.Rev.B}}         
\def\prc{\reff@jnl{Phys.Rev.C}}         
\def\prd{\reff@jnl{Phys.Rev.D}}         
\def\prl{\reff@jnl{Phys.Rev.Lett}}      
\def\pasp{\reff@jnl{PASP}}              
\def\pasj{\reff@jnl{PASJ}}              
\def\qjras{\reff@jnl{QJRAS}}            
\def\skytel{\reff@jnl{S\&T}}            
\def\solphys{\reff@jnl{Solar~Phys.}}    
\def\sovast{\reff@jnl{Soviet~Ast.}}     
\def\ssr{\reff@jnl{Space~Sci.Rev.}}     
\def\zap{\reff@jnl{ZAp}}                        
\def\nat{\reff@jnl{Nature}}             

\def\gsim{\mathrel{\rlap{\lower4pt\hbox{\hskip1pt$\sim$}}
    \raise1pt\hbox{$>$}}}                

\title[Observations of the Corona Borealis supercluster with the 
superextended VSA]{
Observations of the Corona Borealis supercluster with the superextended Very Small Array: 
further constraints on the nature of the non-Gaussian CMB cold spot}

\author[R. G\'enova-Santos et al.] {Ricardo G\'enova-Santos,$^{1,2}\thanks{E-mail:
  rgs@mrao.cam.ac.uk}$ Jos\'e Alberto Rubi\~no-Martin,$^2$ Rafael Rebolo,$^{2,4}$ 
\newauthor Richard A. Battye,$^3$ Francisco Blanco,$^3$ Rod D. Davies,$^3$ Richard J. Davis,$^3$ 
\newauthor Thomas Franzen,$^1$ Keith Grainge,$^1$ Michael P. Hobson,$^1$ Anthony Lasenby,$^1$
\newauthor Carmen P. Padilla-Torres,$^2$ Guy G. Pooley,$^1$ Richard D.E. Saunders,$^1$ 
\newauthor Anna Scaife,$^1$ Paul F. Scott,$^1$ David Titterington,$^1$ Marco Tucci$^2$ and 
\newauthor Robert A. Watson$^3$\\
$^1$ Astrophysics Group, Cavendish Laboratory, University of Cambridge CB3 OHE, UK \\
$^2$ Instituto de Astrofis\'{i}ca de Canarias, 38200 La Laguna, Tenerife, Canary Islands, Spain \\
$^3$ Jodrell Bank Centre for Astrophysics, University of Manchester, Manchester M13 9PL, UK \\ 
$^4$ Consejo Superior de Investigaciones Cient\'{\i}ficas, Spain \\}
\date{Accepted Received In original form}
\pagerange{\pageref{firstpage}--\pageref{lastpage}}
\pubyear{2007}

\begin{document}

\label{firstpage}
\maketitle

\begin{abstract}
We present interferometric imaging at 33~GHz, with the new superextended configuration of
the Very Small Array (VSA), of a very deep decrement in the cosmic microwave
background (CMB) temperature. This decrement is located in the direction of the Corona
Borealis supercluster, at a position with no known galaxy clusters, and was discovered by a
previous VSA survey (G\'enova-Santos et al.). 
A total area of 3~deg$^2$ has now been
imaged, with an angular resolution of 7~arcmin and a flux sensitivity of 5~mJy~beam$^{-1}$.

These observations confirm the presence of this strong and resolved negative spot at
$-41\pm 5$~mJy~beam$^{-1}$ ($-258\pm 29~\mu$K), with a signal to noise level of 8.
This structure is also present in the WMAP 5-year data. The temperature of 
the W-band (94~GHz) data at the position of the decrement agrees within $1.2\sigma_{\rm n}$ with 
that observed by the VSA at 33~GHz, and
within $0.2\sigma_{\rm n}$ with the Sunyaev-Zel'dovich (SZ) spectrum.

Our analyses show that it is a 
non-Gaussian feature in the CMB at a level of 4.8$\sigma$. The probability of finding 
such a deviation or larger in CMB Gaussian simulations is only 0.19 per cent.
Therefore, an explanation other than primordial CMB is required. 
We have considered the possibility of an SZ effect generated in a diffuse, 
extended warm/hot gas distribution. This hypothesis is especially relevant, as
the presence of such structures, if confirmed, could 
provide the location for a significant fraction of the missing baryons in the Local
Universe.
\end{abstract}


\section{INTRODUCTION}

The baryon density at $z=0$, derived from the total budget over the well-observed components
\citep{fukugita_98}, is a factor $\approx$2 lower than that at high redshift, inferred through 
independent methods, namely big bang nucleosynthesis 
\citep{burles_01}, the Ly$\alpha$ forest \citep{rauch_97} and the cosmic microwave background 
(CMB) primary anisotropies (e.g. Rebolo et al. 2004; Dunkley et al. 2008). According to the
results of hydrodynamical simulations \citep{cen_99,dave_01,cen_06} a substantial fraction of these 
missing baryons could be located in the so-called `warm/hot intergalactic medium' (WHIM). 
This is a very diffuse gas phase, arranged in sheet-like or filamentary structures, with 
temperatures $10^5\le T \le 10^7$~K, typical baryon overdensities in the range 
$\delta\rho_{\rm B}/\langle\rho_{\rm B}\rangle\sim 10-30$, and length scales of the order of
10~Mpc. Its intermediate temperature and low density,
in conjunction with the presence of many Galactic foregrounds and different extragalactic
contaminants, make the detection of the WHIM rather challenging. Several attempts 
(some of which have produced tentative detections) have been carried out, 
either by looking for its possible soft X-ray emission
\citep{soltan_02,finoguenov_03,zappacosta_05} or by identifying UV 
\citep{nicastro_02,nicastro_05} or soft X-ray \citep{barcons_05} absorption lines in the
spectra of more distant sources.

The thermal Sunyaev--Zel'dovich (SZ) effect \citep{sunyaev_72} has been proposed, and in fact used,
as an alternative tool to search for this hidden matter. 
This is a secondary anisotropy of the CMB due to the inverse
Compton scattering of its photons by high-energy electrons such as those located in the
extended atmospheres of hot gas ($k_{\rm B} T_{\rm e}\sim 10$~keV) of the richest clusters 
of galaxies. The SZ signal is proportional to the line-of-sight integral of the electron density
multiplied by the electron temperature. This effect is well known and many detections have 
been achieved over the last 
decade in nearby clusters (see e.g. Birkinshaw 1999; Carlstrom, Holder
\& Reese 2002). Structures like
superclusters of galaxies, where the WHIM is likely to be located, could also build up a
significant SZ effect \citep{birkinshaw_99}. Their lower baryon overdensities are compensated
for by 
the long path-lengths of the CMB photons across them. 
Indeed, \citet{monteagudo_06} concluded from the results of a numerical simulation 
that $\approx 15\%$ of the SZ signal on the sky would be generated in this kind of
structure.
Many studies have been carried out to
identify statistical detections of diffuse intra-supercluster (ISC) gas through its likely 
SZ effect, making use of the \textit{COBE}-DMR \citep{banday_96} or WMAP 
\citep{fosalba_03,monteagudo_04,myers_04,monteagudo_04b} datasets. However, there have 
been no clear detections to date.

In this context, we selected the Corona Borealis supercluster (CrB-SC) for observations
at 33~GHz with the VSA extended configuration (G{\'e}nova-Santos et al. 2005, hereafter GS05). 
These were the first targeted 
observations searching for extended SZ in the direction of this kind of structure. We found 
a clear negative feature, which we called ``decrement H'' with a flux density 
$-103\pm10$~mJy~beam$^{-1}$ ($-230\pm 23~\mu$K) and
coordinates RA$=15^{\rm h}22^{\rm m}11.47^{\rm s}$, Dec.$=+28\degr 54'06.2''$ (J2000), 
in a region
with no known clusters. Our analyses ruled out, at a level of 99.6\%, the explanation of 
this decrement based only on a primary CMB anisotropy. This led us to consider an extended SZ
effect generated by a filamentary structure consisting of warm/hot gas. The possibility 
of an SZ effect from an unknown background cluster was also considered, but this is rather 
unlikely due to the large angular size of the decrement, which could only be generated by 
a nearby cluster. The
absence of significant X-ray emission in this region in the {\it ROSAT} XRT/PSPC All-Sky 
Survey map set constraints on the physical characteristics of the hypothetical filamentary 
structure,
making it very unlikely that it could generate such a big SZ decrement on its own. 
Therefore, the most plausible hypothesis is a combination of a negative primary anisotropy and
an extended SZ effect.

This work was followed by an observational campaign in this region with the MITO telescope at 
143, 214, 272 and 353~GHz \citep{battistelli_06}. The results of the analyses of the three
lowest frequency channels, in conjunction with the VSA map at 33~GHz, support the hypothesis
of a combination of a primary CMB anisotropy with an extended SZ effect, the relative 
contribution of the latter component being $0.25^{+0.21}_{-0.18}$. The decrement seems 
to be unresolved by the MITO beam (FWHM$\approx$16'), and is detected in the three 
lowest frequency channels.

In the present work we have carried out observations in the region of this decrement with the
newly superextended configuration of the VSA, which has a factor $\approx$2 finer angular
resolution than the extended configuration used in the previous observations.

In Section~2 we present a description of the VSA interferometer, emphasizing the differences
between the superextended and the previous configurations. In Section~3 we describe the
observations and the data reduction procedure and in Section~4 we present the final maps. 
Section~5 includes the discussion about the possible origins of the decrement, and 
conclusions are presented in Section~6.

\section{THE SUPEREXTENDED VSA}

The VSA is a 14-element heterodyne interferometer, tunable between 26 and 36~GHz with a
bandwidth of 1.5~GHz, sited at the Teide Observatory in Tenerife, at an altitude of 2400~m.
Currently it observes at a central frequency of 33~GHz, with a system temperature of
approximately 25~K. A detailed description of the
instrument can be found in \citet{VSApaperI}. Its main target has been the
observations of the primary anisotropies of the CMB, aimed at estimating its power
spectrum \citep{VSApaperIII,VSApaperV,VSApaperVII}, although it has also been successfully
used to observe the SZ effect in nearby clusters of galaxies \citep{lancaster_05}. 

The VSA has observed with three different sets of antennas. The compact
array, used from October 2000 to September 2001, was sensitive in the multipole
range $\ell\sim 150-900$ and had antenna apertures providing a primary
beam of $4.\degr 6$ FWHM. In October 2001 this was replaced with the 
extended array, sensitive in the multipole range 
$\ell\sim 300-1500$ and with a $2.\degr 1$ FWHM primary beam. 
In 2005 the antenna centres were repositioned and the mirrors replaced to provide a 
$1.\degr 2$ FWHM primary beam. 
This superextended array covers the multipole range
$\ell\sim 350-2450$, and produces a synthesized beam of $\approx$7~arcmin FWHM.
Currently it is chiefly dedicated to the observations of primordial
fields with the objective of investigating the power excess discovered by CBI at
$\ell=2000-4000$ \citep{mason_03}.

Situated next to the main array are the two 3.7-m source-subtractor (SS) antennas for 
simultaneous monitoring of radio sources, providing fluxes at 33~GHz on these sources, 
which are then subtracted from the main array data. This is a two-element interferometer
giving a resolution of 4~arcmin in a 9~arcmin field.

\section{OBSERVATIONS AND DATA REDUCTION}

\subsection{Observations}

The observations were carried out mainly during the periods: i) October and November 2005;
and ii) May and June 2006. We defined 6 pointings, which are listed in
Table~\ref{tab:observations} along with their coordinates. The field CrB-H has the same
coordinates as the corresponding pointing in the previous observations with the extended
configuration (GS05). It was in this field that the decrement H was detected at the highest level 
of significance. The pointing CrB-H-spot is centred on the position where decrement H was found 
in those observations. 
With the aim of producing a
final larger mosaic, we defined four additional pointings 
with coordinates shifted in right ascension and declination.
In Table~\ref{tab:observations} we also quote both total observation and integration times
(the latter indicating the amount of data retained after flagging), and the thermal noise achieved in
each field. Most of the observation time was spent on the central fields CrB-H and
CrB-H-spot. The fractional amount of useful data in the fields CrB-H and
CrB-H-spot exceeds 50\%, but in the four additional fields it is considerably lower. This
is due to the fact that the latter four fields were observed at higher hour angles, where the
quality of the data is worse, and therefore a larger amount of data had to be discarded.
The noises are consistent with the total integration time in each field.
\begin{table*}
\begin{minipage}{150mm}
\begin{center}
\caption{Summary of the observations with the VSA. We list the
central coordinate of the 6 pointings, the total observation time, the
integration time and the achieved thermal noise 
(computed from the map far outside the primary beam).}
\begin{tabular}{lccccc}
\hline
Pointing & RA (J2000) & Dec. (J2000)&$\rm T_{obs}$&$\rm T_{int}$ & Thermal noise \\
   &            &            & (hr)  & (hr) & (mJy~beam$^{-1}$)      \\
\hline
CrB-H         & 15 23 00.00 & 29 13 30.0 &  50 & 33 &  6.8 \\
CrB-H-spot    & 15 22 11.47 & 28 54 06.2 & 145 & 75 &  4.2 \\
CrB-H-spotN   & 15 22 11.47 & 29 24 06.2 &  25 &  8 & 18.3 \\
CrB-H-spotS   & 15 22 11.47 & 28 24 06.2 &  25 & 11 & 15.0 \\
CrB-H-spotE   & 15 23 56.47 & 28 54 06.2 &  15 &  6 & 17.9 \\
CrB-H-spotW   & 15 20 26.47 & 28 54 06.2 &  15 &  6 & 23.2 \\
\hline
\end{tabular}
\label{tab:observations}
\end{center}
\end{minipage}
\end{table*}

\subsection{Calibration and data reduction}

The absolute flux calibration of VSA data is determined from observations of Jupiter, whose
brightness temperature is taken from the WMAP 5-year data: 
$T_{\rm Jup}=146.6\pm 0.7$~K \citep{hill_08}. This flux scale is transferred to other
calibration sources: Tau A and Cas A. A detailed description of the VSA
calibration process is presented in \citet{VSApaperI}. The specifications for the
superextended configuration (such as the correction for the fact that Tau A and Cas A are
partially resolved in the longest baselines) are essentially the same as those adopted for
the extended configuration, and are explained in \citet{VSApaperVII}. As a further
improvement to these processes for the present configuration, a secondary phase
calibration provided by short calibration observations (on Cas A in the observations
presented in this work), 
interleaved with the main field observations, was introduced.

The data reduction pipeline followed essentially the one described in
\citet{VSApaperVII}.
However, in the superextended configuration data some further systematics 
have been identified. Some of the daily maps showed stripes. This is due
to the low quality of the data on the shortest baselines at extreme hour
angles (around $\pm$2${\rm h}$), detected only when stacking the whole dataset in hour angle,
but not on the individual days. By flagging relevant hour angle intervals in these
baselines the stripes are removed. 

\subsection{Source subtraction}

The source subtraction was carried out similarly to GS05. 
This involves extrapolation to 33~GHz of the fluxes of all the sources found in the 
NVSS--1.4~GHz \citep{condon_98} and GB6--4.85~GHz \citep{gregory_96} catalogues. 
In the present observations, due to the narrower primary and synthesized beams, 
we focus on sources closer than $1.2\degr$ from the pointing centres and brighter 
than 18~mJy at 33~GHz. We found that all the
identified sources had measured fluxes above this level, as quoted in 
Table~\ref{tab:sources}. These values
were subtracted from the data, and were obtained, using the same estimator
as in GS05, from the SS observations performed simultaneously with the extended
configuration observational campaign (during the periods May 2003--February 2004 and 
August 2004--November 2004).
\begin{table*}
\begin{minipage}{150mm}
\begin{center}
 \caption{List of all the radio sources with measured fluxes at $33$~GHz above $18$~mJy.
 We quote the extrapolated flux at $33~$GHz, and the estimated flux at $33~$GHz using
 the SS observations, which have been used to subtract the sources from the data. In the last
 three columns we list the number of individual observations for each source, the 
 fractional variation (per cent) of the fluxes on these individual observations and the 
 result of the $\chi^2$ test.}
 \label{tab:sources}
  \begin{tabular}{|c|c|c|c|c|c|c|c|c|}
   \hline
Name  & RA      & Dec.     & $S_{\rm ext}$ & $S_{\rm meas}$ & N$_{\rm obs}$ & $\Delta S/S$ & $\chi^2$ \\
      & (J2000) & (J2000)  &    (mJy)      &   (mJy)        &               &  (\%)     &         \\
     \hline
1516+2918 &  15 16 40.9 &   29 18 25.0  &    37 &   $37\pm 7$ &  16 & 50.4 &   6.0  \\
1519+2746 &  15 19 51.0 &   27 46 23.0  &    33 &   $34\pm 16$&   4 & 82.1 &   3.1  \\
1521+2830 &  15 21 04.8 &   28 30 29.0  &    87 &   $34\pm 4$ &  32 & 62.5 &  36.3  \\
1522+2808 &  15 22 48.9 &   28 08 51.0  &   135 &   $97\pm 3$ & 118 & 40.0 & 172.3  \\
1523+2836 &  15 23 27.9 &   28 36 51.0  &    62 &   $32\pm 3$ &  61 & 78.9 &  45.8  \\
1524+2900 &  15 24 04.2 &   29 00 22.0  &    57 &   $32\pm 3$ &  76 & 65.2 &  45.7  \\
1527+2855 &  15 27 44.7 &   28 55 24.0  &    38 &   $35\pm 7$ &  17 & 68.8 &  12.1  \\
\hline
\end{tabular}
\end{center}
\end{minipage}
\end{table*}
%
%
\begin{figure*}
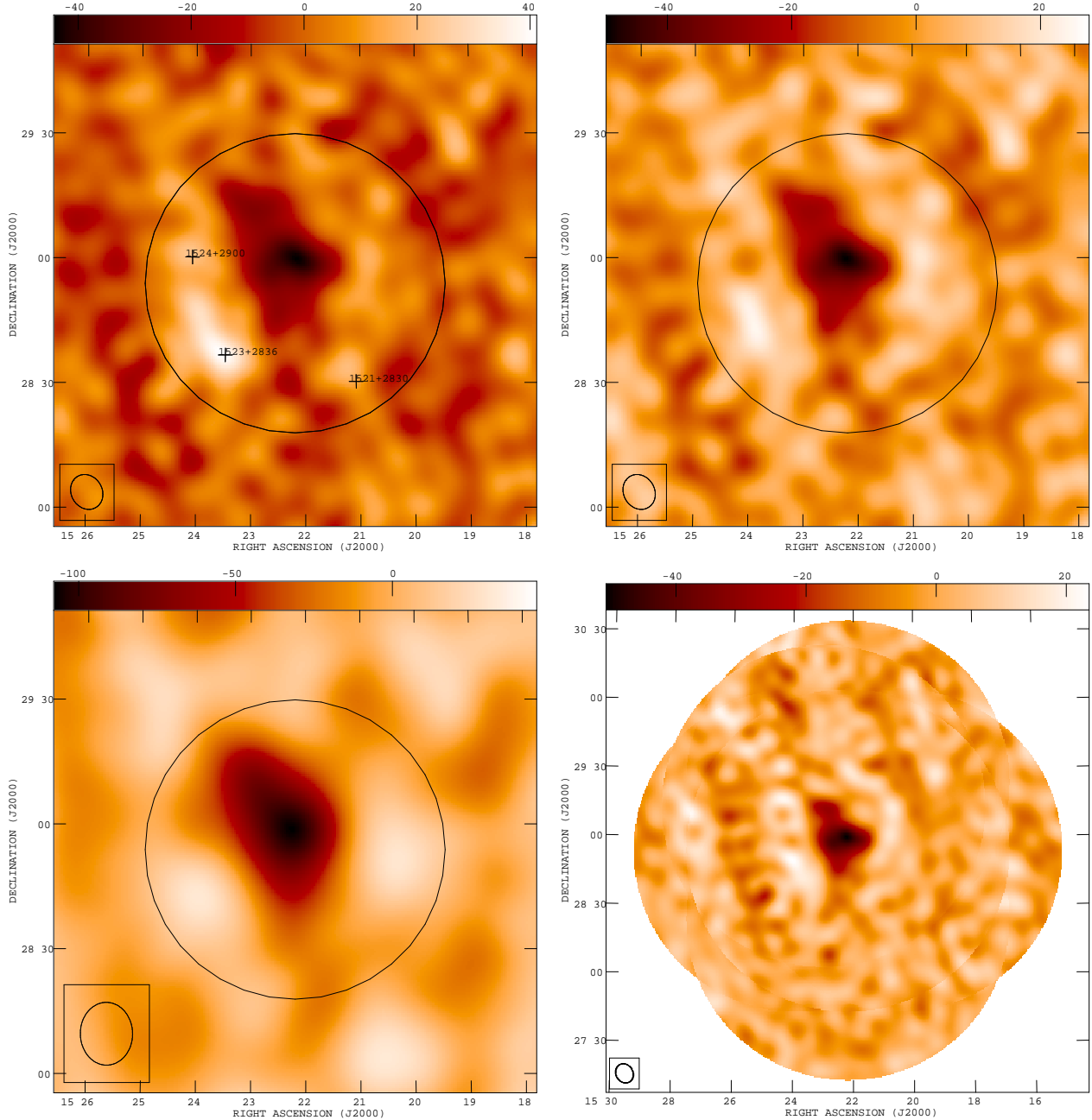

\includegraphics[width=8.5cm]{crbh_spot_vis_final_cln_plt_sour_0.6deg_25.ps}%
\includegraphics[width=8.5cm]{crbh_spot_vis_final_ss_gt18_0.6deg_cln_25.ps}
\includegraphics[width=8.5cm]{crbh_spot_vis_final_ss_gt18_0.6deg_t150lambda_cln_25.ps}%
\includegraphics[width=8.5cm]{crbh_mosaic_ss_gt18_1.2deg_cln.ps}
\caption{\textsc{clean}ed maps of the decrement H in the CrB-SC obtained with the VSA superextended
configuration. Top: maps of pointing CrB-H-spot before (left) and after (right) source subtraction. 
In the non-source-subtracted map the crosses show the positions of the monitored radio sources 
inside the primary beam FWHM. Bottom left:
source-subtracted map of pointing CrB-H-spot, after applying a Gaussian taper of $\sigma=150\lambda$ in the
\textit{uv} plane. 
The circle indicates the primary beam FWHM ($1.2\degr$) of the CrB-Hspot pointing. 
Bottom right: \textsc{clean}ed VSA mosaic built up from the six pointings listed in
Table~\ref{tab:observations}. The noise level is practically uniform across the mosaic at a level $\approx
5$~mJy~beam$^{-1}$. The units of the colour-scale shown in the top bar are mJy~beam$^{-1}$. 
The synthesized beam FWHM is shown in the bottom-left corner of each map.}
\label{fig:maps}
\end{figure*}
\begin{figure*}
\includegraphics[width=6.5cm]{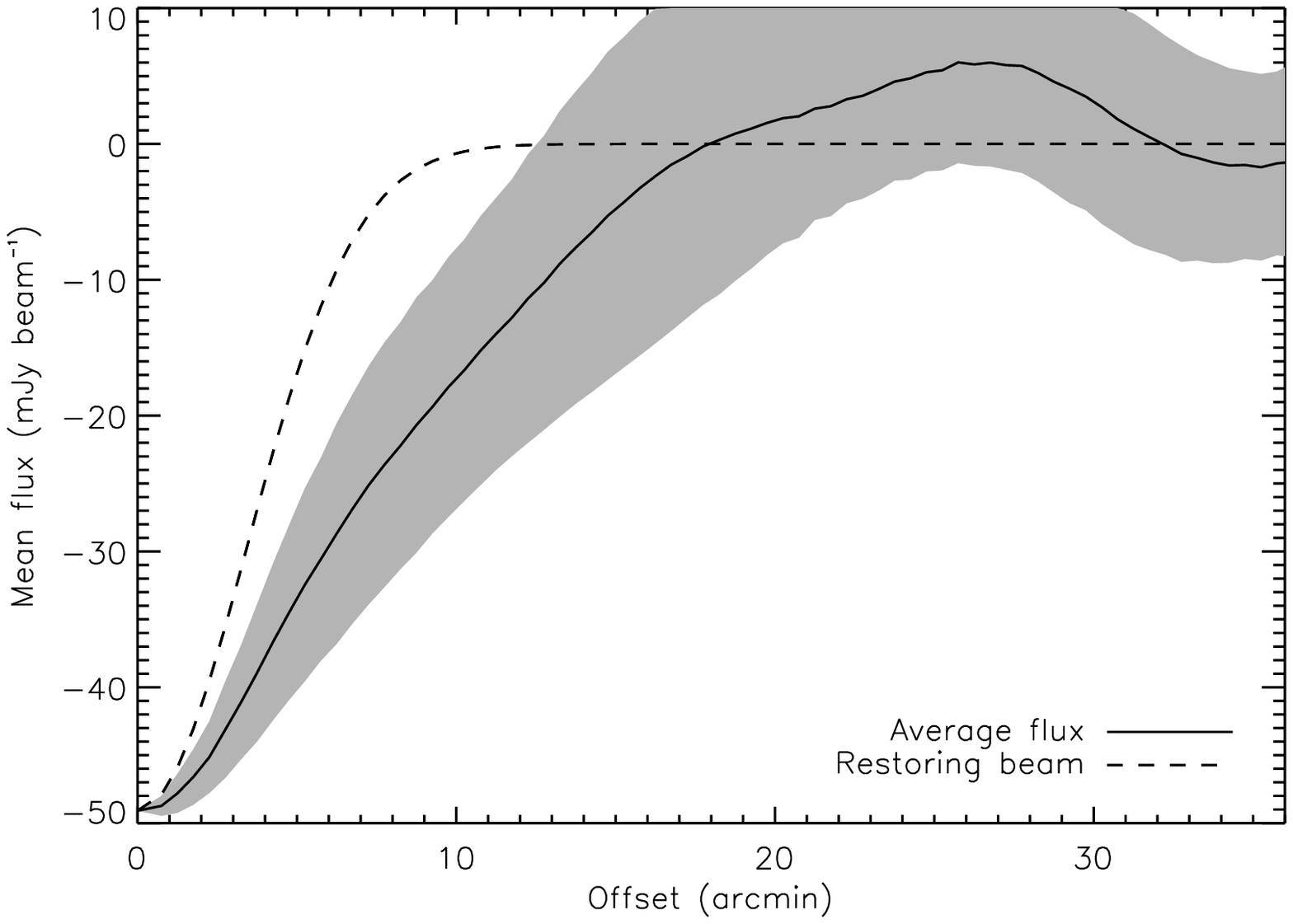}%
\includegraphics[width=6.5cm]{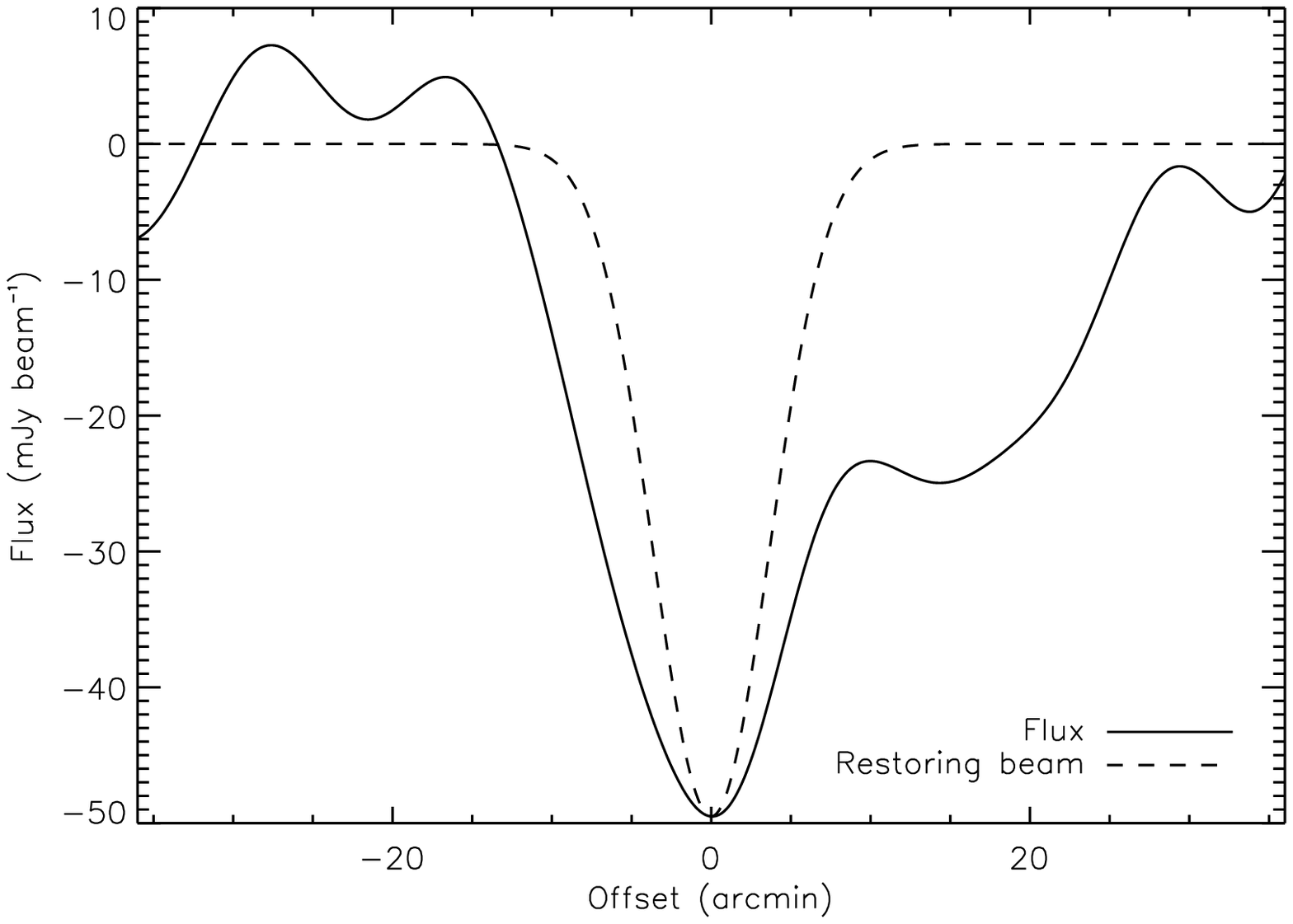}
\includegraphics[width=6.5cm]{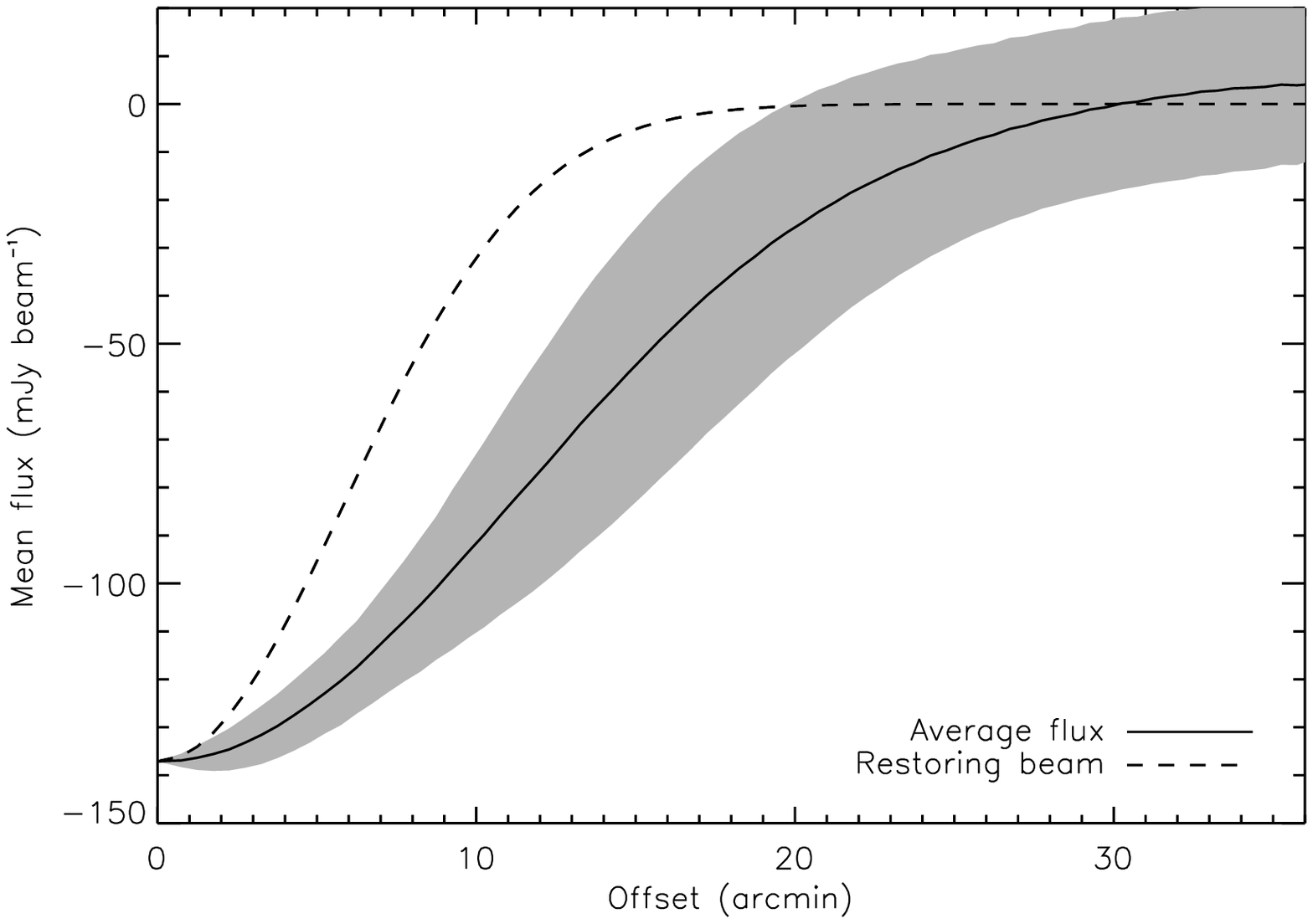}%
\includegraphics[width=6.5cm]{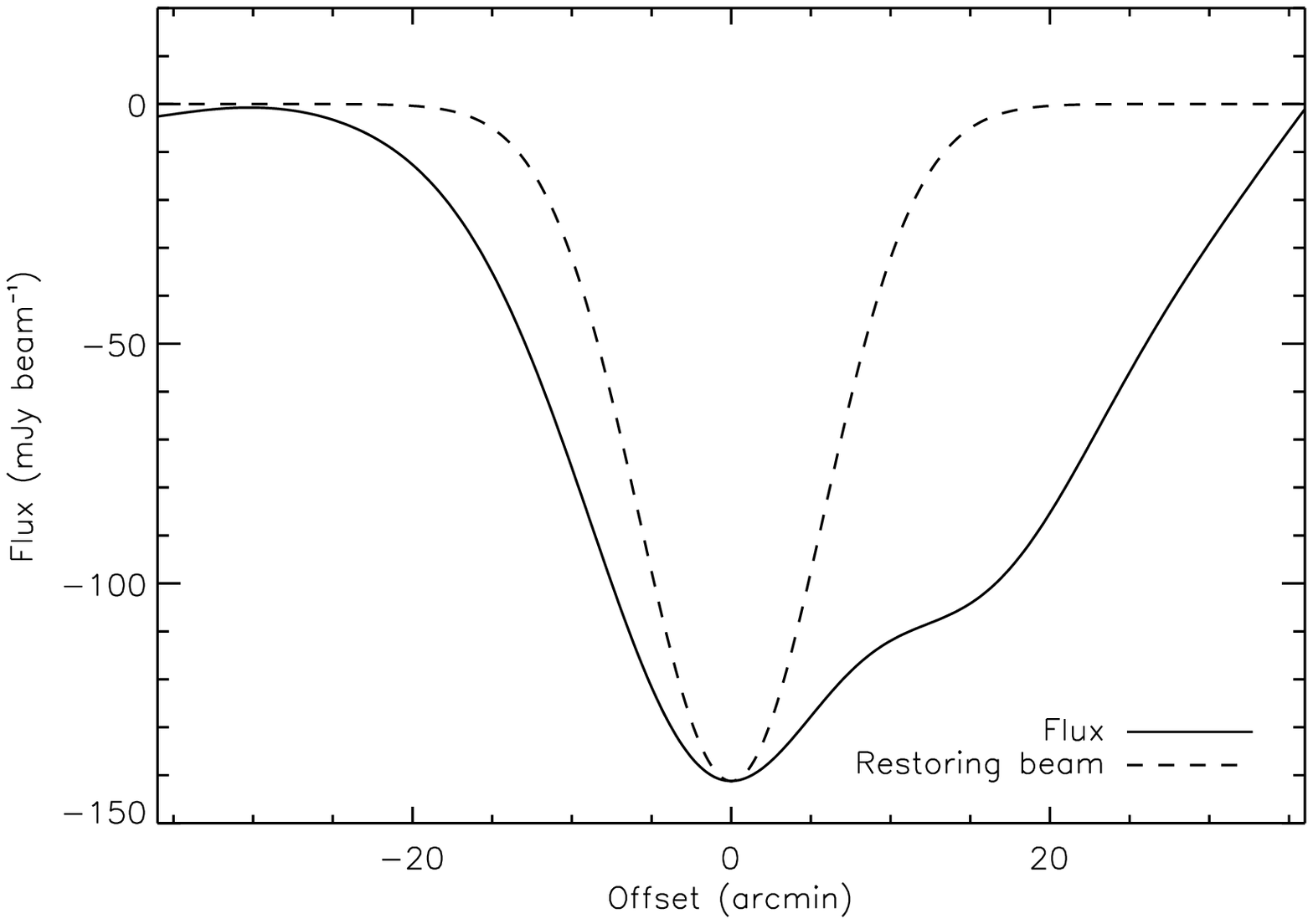}
\caption{Angular profiles of the decrement H in the \textsc{clean}ed and source-subtracted maps of pointing CrB-H-spot
(solid line) and of the restoring beam (dashed line). We represent cases when no tapering
is used (top) and when a Gaussian taper of $\sigma=150\lambda$ is applied to the data (bottom). Left plots show
the average flux in rings of increasing radius from the position of the minimum flux of the decrement, 
whereas right plots represent the flux along a NE-SW line. In the left plots the
shaded regions show the $1\sigma$ error in each ring.}
\label{fig:profiles}
\end{figure*}

This strategy has two main drawbacks. First, as the detection limit of the GB6 
catalogue is 18~mJy, we are missing all the sources with inverted spectra between 4.85~GHz and 
33~GHz, which may be of the order of 20\% of all the sources \citep{waldram_03}. This implies
that we could be missing around one radio source due to this.  
Second, we are subtracting fluxes that have been 
derived from SS observations carried out two years before the observations of the main array. 
Therefore, variability can be important here, and to assess it we have applied a 
$\chi^2$-test to each source.
The values of $\chi^2$ quoted in
Table~$\ref{tab:sources}$ have been calculated as  
\begin{equation}
\chi^2=\sum_{i=1}^{\rm N_{\rm obs}} \left(\frac{S_i-S_{\rm meas}}{\Delta S_i}\right)^2 ~~,
\label{eq:chi2_test}
\end{equation}
where N$_{\rm obs}$ is the total number of observations for each source, 
$S_i$ and $\Delta S_i$ are the value and the error of each individual measurement and
$S_{\rm meas}$ is the final averaged value. This tests the hypothesis that the flux-density 
time-series could be modelled as a constant. Following \citet{bolton_06}, we considered as 
variable the sources for which the probability of obtaining an equal or higher value in a  
$\chi^2$ distribution with N$_{\rm obs}-1$ degrees of freedom is less than 1 per cent.
According to this only the source 1522+2808 is classified as variable. 
Its fractional variation, calculated as 
\begin{equation}
\frac{\Delta S}{S_{\rm meas}}=\frac{1}{S_{\rm meas}}~\sqrt{\frac{\sum S_i^2/(\Delta S_i)^2}{\sum 1/(\Delta S_i)^2}-
\left(\frac{\sum S_i/(\Delta S_i)^2}{\sum 1/(\Delta S_i)^2}\right)^2}~~,
\label{eq:frac_variation}
\end{equation}
is 40\%. However, this source lies at a distance of $0.77\degr$ from the CrB-H-spot pointing 
(this is the most important pointing in our analysis, as will become clear), where the primary 
beam attenuation is a factor 0.32, so the consequences of its variability are 
ameliorated. To check this we have subtracted this radio source from the observed map using 
the two extreme fluxes given by the estimated flux plus and minus its variability, 
and found that the flux difference in the centre of the map is only 1.3\%.

We have estimated the confusion noise due to unresolved sources below our source subtraction
threshold of 18~mJy by using the formalism of \citet{scheuer_57} and the
best-fitting power-law model obtained from the source count of the whole sample of sources in
GS05. We found $\sigma_{\rm sour}=3.8$~mJy~beam$^{-1}$ -- this value is lower
than the thermal noise in any of the fields.

\section{VSA MAPS}

Daily observations, after being calibrated and reduced, are held as visibility files. These
files are loaded individually into \textsc{aips}, where the final stack for each field is
made. Source subtraction is then applied in the aperture plane, and maps are produced
for each pointing using standard \textsc{aips} tasks. Since we are interested in extended 
structures, we typically use natural weighting. The
deconvolution of the synthesized beam is carried out using standard \textsc{aips} tasks  
down to a depth about the noise level.
\textsc{clean} boxes are placed around the position of decrement H, and around the radio 
sources in the non-source-subtracted maps.

The \textsc{clean}ed maps built up from pointing CrB-H-spot, before and after source
subtraction, are shown in the top panels of Fig.~\ref{fig:maps}. Decrement H is clearly
detected (signal-to-noise level of 8), with a peak negative flux density 
$-41\pm 5$~mJy~beam$^{-1}$ (with a $7.7\times 6.5$~arcmin$^2$ beam; 
brightness temperature of $-258\pm 29~\mu$K) at 
RA$=15^{\rm h}22^{\rm m}11.47^{\rm s}$, Dec.$=+29\degr 00'06.2''$ (J2000)
\footnote{This minimum flux value and coordinates (and also other fluxes and coordinates presented 
hereafter) have been obtained from the dirty map. 
This is reliable as the {\sc clean}ing procedure only affects the regions of the map surrounding 
the decrement, where the {\sc  clean} box has been positioned. Therefore, the minimum flux 
of the decrement may remain the same after the {\sc clean}ing.}, and with 
angular size of about $30\times 20$~arcmin$^2$.

In order to enhance the large angular scales, we have applied a Gaussian taper function of
width $\sigma=150\lambda$ to the visibilities. In the resulting map, which is shown in the 
bottom-left corner of Fig.~\ref{fig:maps}, the detection of decrement H is even more
remarkable, with a minimum flux density of $-101\pm 18$~mJy~beam$^{-1}$ (with a 
$15.3\times 10.5$~arcmin$^2$ beam; brightness temperature of $-197\pm 35~\mu$K). 
At this resolution, around the peak decrement there are $\approx 2$ contiguous resolution elements 
each with negative flux exceeding $5\sigma_{\rm n}$ (being $\sigma_{\rm n}$ the thermal noise of 
the map), corresponding to an integrated decrement flux of $-180\pm 19$~mJy; another measure 
of the overall decrement is that there are $\approx 4$ contiguous resolution elements with 
negative flux exceeding $3\sigma_{\rm n}$, corresponding to an integrated decrement of
$-277\pm 26$~mJy.

We have also combined the individual overlapping maps of the six independent pointings to
produce the final source-subtracted mosaic shown in the bottom-right corner of Fig.~\ref{fig:maps}. In order
to avoid the loss of signal-to-noise ratio, and given that the synthesized beams of the six
pointings have similar shapes and orientations, here we have applied the \textsc{clean}ing 
to the whole mosaic instead of to the individual pointings. To this end, we used the
synthesized beam of the pointing CrB-H-spot and placed a \textsc{clean} box around the 
decrement.

In all these maps the decrement is resolved by the VSA synthesized beam. This can be
clearly seen in Fig.~\ref{fig:profiles}, where we plot the angular profiles of both the
decrement and the restoring beam. We have plotted both the radial-averaged profiles (left)
and the profiles along a line oriented in northeast-southwest direction (right), where 
the decrement shows the greatest elongation. It can be seen that there is some
substructure, with a secondary minimum towards the northeast, at a position 
RA$=15^{\rm h}22^{\rm m}36.64^{\rm s}$, Dec.$=+29\degr 04'06.0''$ (J2000).

\begin{figure}
\includegraphics[width=\columnwidth]{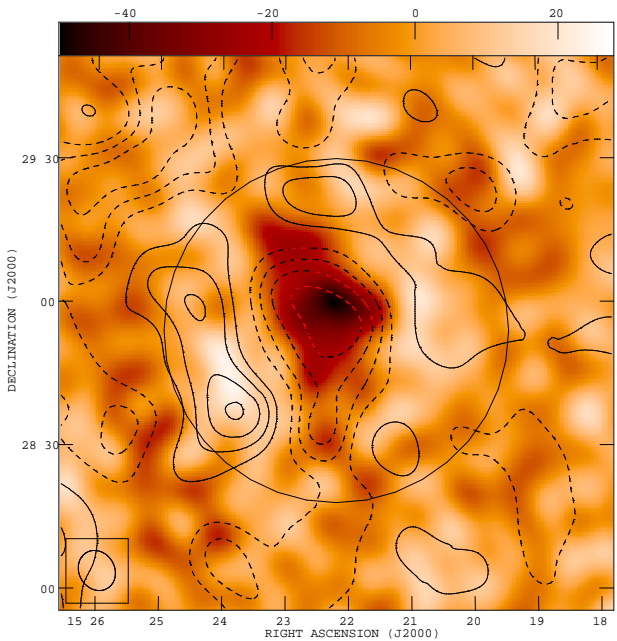}%
\caption{Comparison of the source-subtracted and {\sc clean}ed maps of decrement H obtained with the 
VSA extended configuration (contours) and with the VSA superextended configuration (colour scale). The units of
the superextended map, shown in the top bar, are mJy~beam$^{-1}$. The difference between contours in the extended
map is 18~mJy~beam$^{-1}$ ($1.5\sigma_{\rm n}$). The circle around the centre and the ellipse in the 
bottom-left corner indicate respectively the FWHM of the primary and synthesized beams of the superextended map.}
\label{fig:map_ext_se}
\end{figure}

These results represent a successful confirmation, from a completely
independent dataset, of the previous detection with the VSA extended configuration. In Fig.~\ref{fig:map_ext_se} we
show a comparison of the maps in the region of the decrement resulting from both observational campaigns. 
Despite that in both cases different angular scales are sampled in the aperture plane of our observations, 
the two maps are consistent, and the shape of the structure is similar in them. 
In order to assess the compatibility of the measurements of the minimum brightness temperature
in both observations, 
we have applied a Gaussian taper function of width $\sigma=170\lambda$ to the superextended visibilities. 
This taper function produces a synthesized beam comparable to that of the extended observations. The minimum 
flux decrement in this map is $-90\pm 14$~mJy~beam$^{-1}$ (with a 
$13.9\times 10.3$~arcmin$^2$ beam; brightness temperature of $-196\pm 30~\mu$K), whereas the minimum flux
decrement in the extended map is $-117\pm 15$~mJy~beam$^{-1}$ (with a 
$14.0\times 10.2$~arcmin$^2$ beam; brightness temperature of $-255\pm 33~\mu$K)
\footnote{Note that these fluxes and temperatures have been corrected by the primary beam response, 
which is necessary here as the pixel with the minimum flux lies at different distances from the pointing 
centres in the extended and superextended observations. This is 
why the extended flux and temperature values are different from 
those presented in GS05, where no primary beam correction was applied and the {\sc clean} map was used.}. 
Therefore, both measurements agree to within 1.3$\sigma$. 

%

\section{ORIGIN OF THE DECREMENT}

In GS05 we presented a detailed discussion about the possible origin of decrement H,
considering three different scenarios: a primary CMB anisotropy, an unknown cluster of
galaxies and a concentration of WHIM gas. Our statistical analyses yielded a probability of
0.38\% for a primary anisotropy, while only $\approx$0.3 clusters rich enough to
produce such a deep decrement are expected in the entire surveyed region 
($\approx 24$~deg$^2$). This cluster hypothesis is
further disfavoured by the fact that, due to the absence of any excess of X-ray emission in
the \textit{ROSAT} data, a distant cluster would be needed, and this is in conflict with the
large angular extension of the decrement. Furthermore, we have recently carried out an optical 
survey with the ING telescope that has not shown any evidence of a cluster in the region of 
the CrB-SC (Padilla-Torres et al. 2008, in prep.). A high concentration of infrared galaxies 
at a redshift $z\approx 0.11$ was found, but with a physical distribution unlike that of 
clusters of galaxies.
The absence of X-ray emission also set constraints on the physical characteristics that a
WHIM structure may have.
This indeed rules out the possibility of a diffuse structure, with a depth of the order of 
the maximum separation along the line of sight between clusters in the core of CrB-SC 
($\approx$40~Mpc), 
and with electron temperature typical of
WHIM. Therefore, we concluded that the most plausible hypothesis is a combination of
primordial CMB fluctuations with an SZ effect from a large-filamentary structure oriented in
a direction close to the line of sight. We now revisit these analyses after bringing 
in the new superextended configuration data.

In order to assess the possible contribution from the primordial CMB to this deep
decrement, we repeated the Monte Carlo analysis described in GS05, but now using as template 
in the aperture plane the visibility points sampled by the CrB-H-spot pointing.
We carried out 15000 simulations, adding
to each three components: primordial CMB, thermal noise and residual radio sources below
the subtraction threshold of 18~mJy. For the first of these components, the input power 
spectrum was generated up
to $\ell=3000$ with \textsc{camb} \citep{lewis_00}, from a cosmological model defined by the
following parameters: $\Omega_{\rm B}=0.044$, $\Omega_{\rm M}=0.25$,
$\Omega_{\Lambda}=0.75$, $h=0.73$, $\tau=0.14$, $10^{10}A_{\rm S}=23$, 
$n_{\rm S}=0.97$, as derived from the most recent VSA results \citep{rebolo_04}, plus 
$T_{\rm CMB}=2.725$~K \citep{mather_99} and $\Omega_\nu=0$. For the thermal noise we performed
realizations using the actual noise values in the visibility template. Finally, the residual
radio sources were simulated, in the aperture plane and within $1.2\degr$ of the field centre, 
following the source count derived from the flux values obtained by the SS in the sources 
observed in this region (equation~(1) of GS05).

\begin{figure}
\includegraphics[width=\columnwidth]{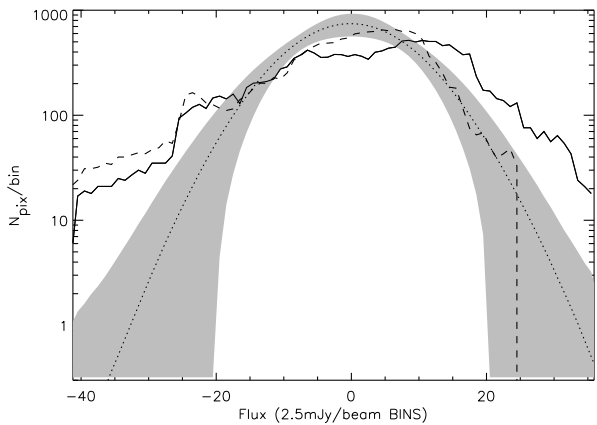}
\caption{Histogram comparing the flux distribution inside the $1.2\degr$ FWHM
of the primary beam in the 15000 simulations (dotted curve) and in the real data of pointing
CrB-H-spot (solid and dashed curves, respectively for the dirty and {\sc clean} maps). We also show 
the $1\sigma$ error bars of the simulations (dashed
region).}
\label{fig:sims_pdf}
\end{figure}

We performed a fluctuation analysis similar to \citet{rubino_03}.
In Fig.~\ref{fig:sims_pdf} we show a logarithmic plot of the histograms, i.e. the $P(D)$
functions, of pixel
values inside the primary beam $1.2\degr$ FWHM from the 15000 simulations (dotted curve) in comparison with
the corresponding distribution of the real data (solid curve). For negative fluxes below $\sim -20$~mJy~beam$^{-1}$
the real data shows a clear excess with regards to the simulations, caused by the
presence of decrement H. In the positive flux tail a similar excess is found, which stems from 
the positive spots around the decrement, especially from the one located towards the southeast 
(see Fig.~\ref{fig:maps}). However, this is an artificial effect, as we
explained in detail in GS05. On one hand, the response of an interferometer has zero mean,
so the presence of a strong negative feature in the map may enhance the neighbouring
positive spots. On the other hand, the convolution of the synthesized beam may
give rise to positive features around the strong negative decrement. Note here that we have 
not applied the deconvolution of the synthesized beam to the simulations, so in 
Fig.~\ref{fig:sims_pdf} we have represented the flux distribution on the dirty maps.
When the flux distribution on the {\sc clean} map is plotted (dashed line) the excess of
the real data in the positive flux tail clearly vanishes, while the excess in the negative
flux tail remains.
\begin{figure*}
\includegraphics[width=12cm]{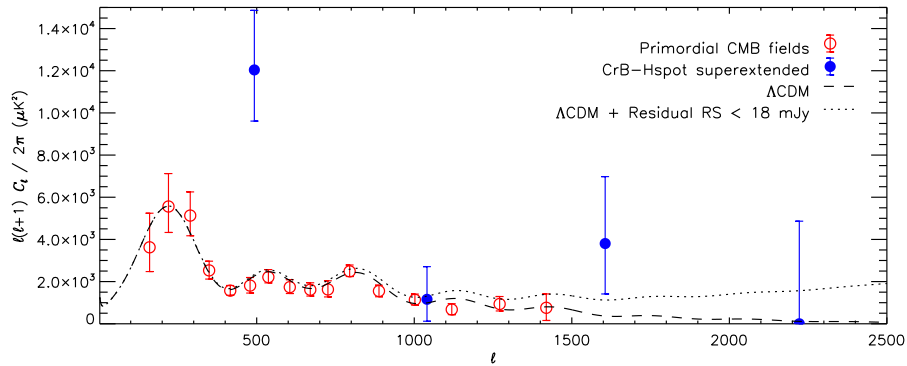}
\caption{Power spectrum computed from the pointing CrB-H-spot (filled circles, blue online), in comparison
with that obtained from the VSA primordial fields observations (Dickinson et al. 2004; open circles, red
online). We also plot a $\Lambda$CDM power spectrum (generated as explained in section~5),
with and without the contribution from residual radio sources below 18~mJy. The strong
deviation at $\ell\approx 500$ is due to the presence of decrement H.} 
\label{fig:ps}
\end{figure*}

The percentage of realizations in which the minimum CMB flux value is below that found in
the real map is 0.19\%. Although a factor of two lower, this value is of the order of 
that obtained when the equivalent study was applied to the extended configuration data (GS05). 
We computed in the simulations the standard deviations of all pixels within the primary beam
FWHM in order to estimate the confusion level introduced by each component, finding:
$\sigma_{\rm CMB}=6.6$~mJy~beam$^{-1}$, $\sigma_{\rm n}=4.8$~mJy~beam$^{-1}$ and 
$\sigma_{\rm sour}=2.5$~mJy~beam$^{-1}$. Adding these in quadrature we obtain 
$\sigma=8.6$~mJy~beam$^{-1}$, from which we conclude that decrement H is a 4.8$\sigma$
deviation.

We have also computed the power spectrum of pointing CrB-H-spot, following the procedure
described in \citet{hobson_02}. The result is shown in Fig.~\ref{fig:ps}, along with that obtained
from the last published primordial CMB observations with the VSA \citep{VSApaperVII} and, for
comparison, the theoretical
power spectrum that we have used in the simulations. In the first bin, at
$\ell\approx 500$, the power spectrum of the CrB-H-spot pointing shows a $4.0\sigma$ 
deviation from the pure primordial CMB behaviour. This is caused by the presence of the decrement,
as this angular scale corresponds to its size, and is similar to the $2.4\sigma$ deviation at
$\ell\approx 550$ obtained from the extended configuration data (GS05).

These analyses clearly highlight the non-Gaussian nature of this decrement, and show that it 
is unlikely to 
be caused by a primordial CMB fluctuation, confirming the results obtained in GS05. 
Therefore, as the possibility
of an unknown galaxy cluster seems implausible (see above), the hypothesis of a large
filament pointing towards us is reinforced. The conclusions about the physical shape, density and
temperature that this hypothetical structure may have are similar to those put forward in GS05. 
Essentially, a short filament located inside the CrB-SC and with an electron temperature within the
typical WHIM range ($0.01-1$~keV), can not account itself for the total SZ 
signal, without any significant X-ray emission. 
However, we may consider that the total decrement could be a combination of a negative
primary anisotropy and an SZ effect. In fact, this decrement is detected in the 214~GHz
channel (close to 218~GHz, the null of the thermal SZ effect) 
of MITO \citep{battistelli_06}, indicating that there might be a primordial
component. These MITO observations yield that a relative fraction $f=0.25$ of the total 
decrement is produced by a thermal SZ effect. However, up to $f\approx 0.7$ could be
produced by this effect within the typical WHIM temperatures and without a significant X-ray
emission. If, for instance, we assume that half ($f=0.5$) of the
decrement is due to the SZ effect from a short filament with electron temperatures
$T_{\rm e}\sim 0.6-0.8$~keV, then baryon overdensities of the order
$\delta\rho_{\rm B}/\langle\rho_{\rm B}\rangle \sim 1000-1200$ would be needed. These
values are in tension with the results of \textit{N}-body galaxy formation simulations, which
predict that most of the WHIM baryons ($70\%-80\%$) lie in baryon overdensities within the range
$5-200$ \citep{dave_01}. However, there is still a significant WHIM fraction at
$\delta\rho_{\rm B}/\langle\rho_{\rm B}\rangle\sim 1000$ (see Fig.~4 of \citet{dave_01}), and
therefore overdensities of this order are also possible. If we consider a larger filament, then
these restrictions are relaxed. We could think of a filament connecting the CrB-SC with the region
at $z\approx 0.11$ where Padilla-Torres et al. 2008 (in prep.) have found a large concentration of
infrared galaxies. It would have a length of $\approx 130$~Mpc, hold a gas mass 
$M_{\rm gas}\approx 10^{15}$~M$_\odot$, and could produce an SZ decrement as deep as the structure
detected in the map ($f=1.0$) without a detectable X-ray emission. However, according to the
simulations, it is difficult to find filaments of this size. This fact, together with the required 
high elongation along the line of sight, make this structure rather unlikely.

\begin{figure*}
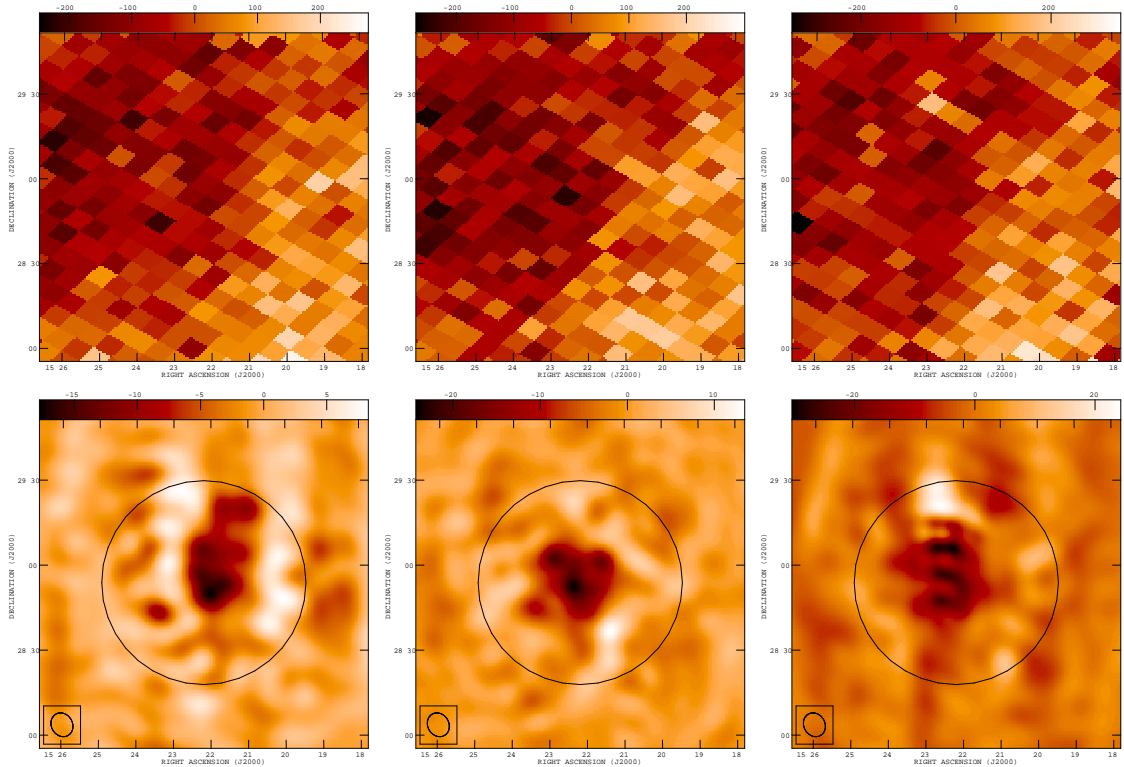

\includegraphics[width=5cm]{wmap_5yr_Q_at_crbh_spot.ps}%
\includegraphics[width=5cm]{wmap_5yr_V_at_crbh_spot.ps}%
\includegraphics[width=5cm]{wmap_5yr_W_at_crbh_spot.ps}
\includegraphics[width=5cm]{wmap_5yr_Q_conv_crbh_spot.ps}%
\includegraphics[width=5cm]{wmap_5yr_V_conv_crbh_spot.ps}%
\includegraphics[width=5cm]{wmap_5yr_W_conv_crbh_spot.ps}
\caption{WMAP 5-year maps \citep{hinshaw_08} in the region of pointing CrB-H-spot.
In the top panels the foreground reduced maps of bands Q, V and W (from left to right) are
depicted. The maps in the bottom panels have been obtained by multiplying and convolving the 
top maps with the VSA primary and synthesized beams, respectively. 
The units of the
colour-scale are $\mu$K in the top panels and mJy~beam$^{-1}$ in the bottom panels.}
\label{fig:maps_wmap}
\end{figure*}

It is worthwhile to examine the WMAP maps in the position of decrement H. In the top
panels of Fig.~\ref{fig:maps_wmap} we show the WMAP 5-year temperature maps
corresponding to the frequency bands Q, V and W \citep{hinshaw_08}, centred on the
coordinates of CrB-H-spot pointing.
The temperature towards the centre of these maps seem to be mostly negative, at a level of 
$\sim -150~\mu$K. However, it is clear that owing to the coarser angular resolution and
lower signal-to-noise of the WMAP data a detailed comparison with our maps is difficult to
perform.
In order to obtain a more realistic comparison we have convolved these maps with the VSA 
synthesized beam. To do this we have multiplied the WMAP maps by the VSA primary beam, 
inverse-Fourier-transformed them to the aperture plane, sampled the visibilities in the 
{\it uv} points of the CrB-H-spot stack, and finally Fourier-transformed them to the map 
plane. As a consequence of this, the resulting maps, which are shown 
in the bottom panels of Fig.~\ref{fig:maps_wmap}, contain only the angular 
scales sampled by the overlapping of WMAP and VSA window functions. 
In these maps the
decrement H appears at a high level of significance, especially in the Q and W frequency
bands.

The VSA {\textit uv}-tapered map shown in the bottom-left corner of 
Fig.~\ref{fig:maps} has a synthesized beam with FWHM$\approx 15.3'\times 10.5'$ (in two
orthogonal directions). This is comparable to the 13.2' FWHM of the WMAP W-band beam,
and indeed both beams subtend the same solid angle. Therefore, a comparison between the 
temperature values derived from both maps is reliable. The minimum flux in the VSA map is
found at position RA$=15^{\rm h}22^{\rm m}20.62^{\rm s}$, Dec.$=+28\degr 59'06.2''$ 
(J2000) and corresponds to a brightness temperature (RJ) of $-197\pm 35~\mu$K.
The thermodynamic temperature in the WMAP map on this position is $-151\pm 66~\mu$K,
which, at the frequency of 94~GHz of the W band, is equivalent to a brightness 
temperature of $-120\pm 53~\mu$K. Therefore, the VSA and
WMAP temperature values agree at a level of $1.2\sigma_{\rm n}$. On the other hand, 
taking into account
that the ratio of the SZ spectrum for the brightness temperature between 33~GHz and 94~GHz
is 
1.51\footnote{This number has been obtained by considering the SZ effect spectrum for the 
brightness temperature,
\[
h(x)=\frac{x^2e^x}{(e^x-1)^2} \left[x\frac{e^x+1}{e^x-1}-4 \right]
\]
with $x=h\nu/(k_{\rm B} T)$, between the WMAP 
($\nu=94$~GHz) and VSA ($\nu=33$~GHz)
frequencies.},
the WMAP value would correspond to a brightness temperature at 33~GHz of $-182\pm
80~\mu$K. Therefore, the VSA and WMAP values are compatible within their error bars with an SZ
effect. 

There are many other mechanisms producing primary or secondary anisotropies which are worth 
discussing here, although they usually lead to CMB fluctuations either weaker or on smaller angular 
scales than those produced by the thermal SZ effect.\\
{\it Kinetic SZ}.
It is well known that in galaxy clusters the kinetic SZ effect is much lower than the thermal
component, typically an order of magnitude smaller. However, at lower electron temperatures, like
those of the WHIM, both components become comparable.  
In a recent work \citet{atrio_08} argued that the kinetic SZ effect generated in
WHIM structures could have an important contribution. They estimate that 1\% of all WHIM
filaments of size $\sim 5$~Mpc, aligned with the line-of-sight, produce a combined thermal and
kinetic SZ effect that could account for the total flux density observed in decrement H. In
particular, a filament of size $L\approx 5$~Mpc and electron density $n_{\rm e}\approx 10^3$~m$^{-3}$
could produce a combined thermal and kinetic SZ signal of $\approx -200~\mu$K. However, a structure
like this should produce an X-ray signal detectable in the \textit{ROSAT} data.\\
{\it Cosmic texture}.
Also recently \citet{cruz_07} argued that the ``cold spot'' detected in WMAP data \citep{vielva_04}
could be caused by a texture, a type of cosmic defect due to symmetry-breaking phase
transitions in the early Universe that may produce positive and negative features on the
CMB \citep{turok_90}. This spot has an angular size of $\approx 10\degr$, and a minimum
thermodynamic temperature of $\approx -550~\mu$K in Q, V and W bands. The predicted number of
textures is proportional to $\theta^{-2}$ ($\theta$ being its angular size), so many more spots
like this are expected in the sky at lower angular scales. Therefore, this phenomenon must be regarded as an
alternative explanation for decrement H. 
Equation~(15) of \citet{turok_90} allow us to estimate the number of textures expected in the
whole sky of a given angular size. Using that formalism we estimate that the number of textures of
size between 10 and 40~arcmin (the angular scale of decrement H) in the whole area initially
surveyed in the CrB-SC (24~deg$^2$) is 0.3. In the total area surveyed by the VSA, combining the
CrB-SC survey with the primordial fields observations (82~deg$^2$; Dickinson et al. 2004), where
no features of the size and amplitude of decrement H have been found, the
total number expected is 1.2. Therefore, it would be likely to find a structure of this size in
the total area observed by the VSA. However, it must be noted that different small-scale
processes such as photon diffusion could smear out textures at sub-degree angular scales.
Current simulations have not enough angular 
resolution to study such small angular scales, so it is not clear whether textures of this size 
could exist.\\
{\it Lensing of the CMB}.
A massive structure like the CrB-SC could lens the CMB photons, leading to distortions on the mean
CMB temperature. However, these distortions are usually small. Even in the richest galaxy clusters, 
CMB lensing has a characteristic amplitude of a few microkelvins, although its angular size can
extend out to a fraction of a degree \citep{seljak_00,holder_04}. In superclusters like
CrB, with much lower densities in the inter-cluster medium, and whose member clusters
are not very rich, even lower signals are expected. Therefore, a temperature fluctuation 
of $\approx -300~\mu$K like
decrement H can not be explained in any way by this effect.\\
{\it Integrated Sachs-Wolfe and Rees-Sciama effects}. When CMB photons cross time-varying
potentials they are blue (in decreasing potentials) or red-shifted (in increasing potentials). This
mechanism is known as the Integrated Sach-Wolfe (ISW) or the Rees-Sciama (RS) effect, 
when the density fluctuation associated with the time-varying potential is evolving in the linear or
non-linear regime, respectively. These effects build up anisotropies in the CMB in the direction of
extremely massive structures or voids, typically at large angular scales ($\gsim 5\degr$) and with
amplitudes of the order of a few 
microkelvins (see, e.g., 
Mart\'{\i}nez-Gonzalez \& Sanz 1990; Puchades et al. 2006). With such angular scales and amplitudes 
this effect is not likely to have a significant contribution to decrement H.
However, \citet{dabrowski_99} studied the physical properties a galaxy cluster may have to produce 
a combined SZ and RS signal which could account for the CMB decrement detected by \citet{jones_97} 
in the direction of the quasar pair PC~1643+4631~A\&B without X-ray emission, and found 
a more significant RS signal. According to their calculations a galaxy cluster at $z=1$ with 
$T_{\rm e}\sim 1.3$~keV and enclosing a total mass of $\sim 10^{16}$~M$_\odot$, would
produce SZ and RS effects respectively of the order $\sim 500~\mu$K and $\sim 250~\mu$K without a
significant X-ray emission. But, as we have remarked, a distant cluster is not a likely 
explanation for decrement H owing to its large angular size.

\section{CONCLUSIONS}

We have carried out observations at 33~GHz with the VSA superextended configuration in the direction 
of a very deep decrement previously detected by observations with the VSA extended configuration in
the CrB-SC (GS05), in a region with no known galaxy clusters and without a clear excess 
of X-ray emission. Our results clearly confirm the 
presence of this strong decrement, 
with flux density of $-41\pm 5$~mJy~beam$^{-1}$ ($-258\pm 29~\mu$K) 
and coordinates RA$=15^{\rm h}22^{\rm m}11.47^{\rm s}$, Dec.$=+29\degr 00'06.2''$ (J2000). It
has an angular size of about $30\times 20$~arcmin$^2$, being clearly resolved by the superextended 
configuration synthesized beam (FWHM$\approx$7~arcmin).

We have confirmed that this deep decrement is detected in the WMAP 5-year maps, but at
a much lower level of significance as a consequence of WMAP's lower angular resolution and 
signal-to-noise. The temperature at the position of the decrement in the W-band map 
agrees at a level of $1.2\sigma_{\rm n}$ with the VSA value, but both temperatures are consistent with a 
SZ spectrum within their error bars.

A study based on simulations of VSA observations has confirmed its non-Gaussian nature. This study 
has shown that it is a $4.8\sigma$ fluctuation, where sigma here includes the confusion noises 
introduced by the primordial Gaussian CMB, thermal noise and residual radio sources. The 
probability of this decrement being entirely produced by a combination of these three 
components is only 0.19\%. Therefore an
alternative explanation needs to be proposed, and we considered the most likely to be the
SZ effect. A galaxy cluster, respectively rich and close enough to produce such a deep and 
large decrement, would probably have already been detected by existing optical or X-ray
surveys. Therefore, this SZ effect could be produced in a less dense and hot structure such 
as the WHIM. In order to produce such a deep decrement without a significant X-ray emission
a filamentary structure with an elongation along the line-of-sight of about 100~Mpc and a 
baryon overdensity of 500-800 would be 
required. These numbers lie close to the top limit of what is expected in these kind of structures,
so a filament like this would be rather unlikely. However, if we consider that the total decrement 
could be a combination of both a primordial CMB fluctuation and an extended SZ effect, then we could
have a structure with a size, density and temperature typical of WHIM producing about the half of the
minimum flux.

This study reinforces the conclusions addressed in GS05. We are confident that our
measurements do show an excess decrement with regard to the primordial CMB. To explain this
decrement we require an SZ effect, probably generated in a large-scale WHIM structure. If 
confirmed, this would have been the first direct detection of an SZ effect from these
structures. This would encourage searches for similar structures in other superclusters by
other CMB experiments. If they are found, then a significant fraction of the missing
baryons in the Local Universe will have been identified.

\bibliography{coronab_paper}

\bibliographystyle{mn2e}


\bsp 

\label{lastpage}

\end{document}